\documentclass[preprint,showpacs,preprintnumbers,amsmath,amssymb]{revtex4}

\usepackage{graphicx}
\usepackage{dcolumn}
\usepackage{bm}

\begin{document}

\title{Fragmentation of relativistic nuclei in peripheral interactions in
 nuclear track emulsion}

\author{D.~A.~Artemenkov}
   \affiliation{Joint Insitute for Nuclear Research, Dubna, Russia}
 \author{V.~Bradnova}
   \affiliation{Joint Insitute for Nuclear Research, Dubna, Russia} 
\author{M.~M.~Chernyavsky}
  \affiliation{Lebedev Institute of Physics, Russian Academy of Sciences, Moscow, Russia} 
\author{L.~A.~Goncharova}
  \affiliation{Lebedev Institute of Physics, Russian Academy of Sciences, Moscow, Russia} 
\author{M.~Haiduc}
   \affiliation{Institute of Space Sciences, Magurele, Romania}
\author{N.~A.~Kachalova}
   \affiliation{Joint Insitute for Nuclear Research, Dubna, Russia} 
\author{S.~P.~Kharlamov}
   \affiliation{Lebedev Institute of Physics, Russian Academy of Sciences, Moscow, Russia}
\author{A.~D.~Kovalenko}
   \affiliation{Joint Insitute for Nuclear Research, Dubna, Russia}  
\author{A.~I.~Malakhov}
   \affiliation{Joint Insitute for Nuclear Research, Dubna, Russia} 
\author{A.~A.~Moiseenko}
   \affiliation{Yerevan Physics Institute, Yerevan, Armenia}
\author{G.~I.~Orlova}
   \affiliation{Lebedev Institute of Physics, Russian Academy of Sciences, Moscow, Russia} 
\author{N.~G.~Peresadko}
   \affiliation{Lebedev Institute of Physics, Russian Academy of Sciences, Moscow, Russia} 
\author{N.~G.~Polukhina}
   \affiliation{Lebedev Institute of Physics, Russian Academy of Sciences, Moscow, Russia} 
\author{P.~A.~Rukoyatkin}
   \affiliation{Joint Insitute for Nuclear Research, Dubna, Russia} 
\author{V.~V.~Rusakova}
   \affiliation{Joint Insitute for Nuclear Research, Dubna, Russia} 
\author{V.~R.~Sarkisyan}
   \affiliation{Yerevan Physics Institute, Yerevan, Armenia}    
\author{R.~Stanoeva}
  \affiliation{Institute for Nuclear Research and Nuclear Energy, Sofia, Bulgaria}
\author{T.~V.~Shchedrina}
   \affiliation{Joint Insitute for Nuclear Research, Dubna, Russia}
 \author{S.~Vok\'al}
   \affiliation{P. J. \u Saf\u arik University, Ko\u sice, Slovak Republic}
\author{A.~Vok\'alov\'a}
   \affiliation{P. J. \u Saf\u arik University, Ko\u sice, Slovak Republic}    
\author{P.~I.~Zarubin}
     \email{zarubin@lhe.jinr.ru}    
     \homepage{http://becquerel.lhe.jinr.ru}
   \affiliation{Joint Insitute for Nuclear Research, Dubna, Russia} 
 \author{I.~G.~Zarubina}
   \affiliation{Joint Insitute for Nuclear Research, Dubna, Russia}

\date{\today}
\begin{abstract}
The technique of nuclear track emulsions is used to explore the 
fragmentation of light relativistic nuclei down to
the most peripheral interactions - nuclear \lq\lq white\rq\rq ~stars. A complete pattern 
of therelativistic dissociation of  a $^8$B nucleus with target fragment 
accompaniment is presented. Relativistic dissociation $^{9}$Be$\rightarrow$2$\alpha$ is explored using significant
statistics  and a relative contribution of $^{8}$Be  decays from 0$^+$ and 2$^+$ states is established. Target
 fragment accompaniments are shown for relativistic fragmentation $^{14}$N$\rightarrow$3He+H and 
$^{22}$Ne$\rightarrow$5He. The leading role of the electromagnetic dissociation on heavy nuclei with respect to 
break-ups on target protons is demonstrated in all these cases. It is possible to conclude that the peripheral dissociation of
 relativistic nuclei in nuclear track emulsion is a unique tool to study many-body systems 
composed of lightest nuclei and nucleons in the 
energy scale relevant for nuclear astrophysics.
\end{abstract}
 \pacs{21.45.+v,~23.60+e,~25.10.+s}

\maketitle
\section{\label{sec:level1}Introduction}
\indent
Nuclear beams of an energy higher than 1~\textit{A}~GeV are recognized as a modern tool used for the study of the structure
of atomic nuclei (a recent review  \cite{Aumann05}). Among the variety of nuclear 
interactions the peripheral dissociation 
 bears a uniquely complete information about the excited states above particle decay thresholds. The peripheral 
dissociation is revealed as a narrow jet of relativistic fragments the summary charge of which is close to the charge 
of the primary nucleus. In spite of the relativistic velocity of motion the internal velocities in the jet  are 
non-relativistic \cite{Zarubin06}. Information about the generation of such fragment ensembles can be useful 
in nuclear astrophysics (indirect approaches), as well as in develoments of
 nucleosynthesis scenarios on the basis of multi-particle fusion.
 It is necessary to provide a complete observation of fragments to utilize 
this possibility.\par

\indent The difficulties of principle are here as follows. An increase in the dissociation degree 
of a relativistic nucleus 
 leads to a decrease in the response of the fragment detector. This circumstance
 makes the complete analysis of 
relativistic fragments, which is necessary up to the He and H isotopes, rather difficult. The excited state is identified
 by the invariant mass of the relativistic fragment jet. Therefore the most accurate measurements of the emission angles of 
fragments are needed. The accuracy of measurements of the momenta is not so rigid, it is possible to assume that the 
fragments conserve the primary momentum per nucleus. In addition, the selection of extremely peripheral  collisions 
requires that the threshold of detection of the target fragments in a total solid angle would be reduced to a minimum. 
\par

\indent
The nuclear emulsion technique solves these problems and makes it possible to perform rather effectively survey investigations
on newly produced beams \cite{Becquerel}. Unique information about the structure of peripheral dissociation of many 
nuclei has already been obtained in \cite{El-Naghy88,Baroni90} and etc. Limitations imposed to statistics are compensated by the fact that the fragment jet structures are 
inaccessible for observation in other approaches.   Besids, the emulsion compound containing
 both the H, Ag, and Br nuclei in comparable concentrations (ratio about 3.2/1./1.) turns out to be use
ful for comparing  interactions. Under the same conditions it is possible to observe the
 very peripheral break-up in the electromagnetic field of a heavy target nucleus as well as in collisions with 
target protons.
\par
\indent
The response of the emulsion nuclei includes the multiplicity of strongly ionizing target fragments from $\alpha$ particles
up to recoil nuclei \textit{n$_b$} and non-relativistic H nuclei \textit{n$_g$}. Besides, the reactions are 
characterized by the multiplicity of produced mesons \textit{n$_s$}. The events in which there are no tracks of target
 nucleus fragmentation belong to dissociation on Ag, Br and are named \lq\lq white\rq\rq ~  stars 
(\textit{n$_b$}~=~0, \textit{n$_g$}~=~0, \textit{n$_s$}~=~0) \cite{Baroni90}. Dissociation on a proton must lead to
 the appearance of its track, that is, \textit{n$_b$}~=~0, \textit{n$_g$}~=~1, and \textit{n$_s$}~=~0. The structure of the events of these two types is just the subject of the
 present paper.
\par
\indent
The presence of strongly ionizing particle (\textit{n$_b>$}~0) tracks in the vertex or relativistic particle (\textit{n$_s>$}~0) tracks outside the 
fragmentation cone makes it possible to define the interaction as the one which is occurred with an overlap of the densities
 of colliding nuclei or with C, N and O nuclei in the cases of extremely short tracks of recoil nuclei. In principle, mutual
 excitation and simultaneous fragmentation of both colliding nuclei are possible. The discussion of these events is outside 
the scope of the present paper and their statistics is given for the sake of an illustration.
\par
\indent
In what follows, the  of the $^8$B, $^9$Be and $^{14}$N dissociation channels are discussed. 
The results are obtained in a BR-2 
emulsion with relativistic sensitivity which was exposed to the JINR nuclotron beams. 
The secondary $^8$B and $^9$Be beams were 
produced in the fragmentation of the primary $^{10}$B beam at an energy about 1.2~\textit{A}~GeV \cite{Rukoyatkin06}.
\par

\section{\label{sec:level2}Fragment  Accompaniment of $^8$B Dissociation and Prospect Studies of $^{9,10}$C  Nuclei}
\indent
Due to the small values of the proton binding energy the  $^8$B nucleus is a sensitive probe of the type of the interaction 
up to the lowest momentum transfers. The study of the events with a total relativistic fragment charge of
$\Sigma$Z$_{fr}$~=~5 in an 
emulsion exposed to  $^8$B nuclei enabled one to establish the leading contribution of the \lq\lq white\rq\rq ~  stars
 $^8$B$\rightarrow^7$B+\textit{p} as compared with the
stars containing the target fragments \cite{Stanoeva06,Artemenkov07}. 
This fact shows a qualitative difference from $^{10}$B nuclei for which \lq\lq white\rq\rq ~  stars 2He+H were predominant 
\cite{Andreeva05}.
\par
\indent
A detailed distribution of the $^8$B dissociation over the fragment configurations  $\Sigma$Z$_{fr}$ and the 
numbers of the target
fragments \textit{n$_b$} and \textit{n$_g$} is given in Table~1. First of all, the predominance of
 \lq\lq white\rq\rq ~  stars $^8$B$\rightarrow^7$Be+\textit{p} should be noted (example is shown in Fig.~1). In this channel, 
there is practically no dissociation on protons \textit{n$_g$~}~=~1. 
The difference is due to a rapid increase in the EM dissociation cross section
 with increasing target nucleus charge (like Z$^2$). 
Half a number of \lq\lq white\rq\rq ~  stars is associated  with 3- and 4- particle 
dissociation modes having much higher thresholds. This implies that the multiple fragmentation 
can be initiated by an EM excitation.
It may also be noted that in the 2He+H and He+3H channels the fraction of the events on protons (\textit{n$_g$}~=~1) 
and the events with 
target fragments (\textit{n$_b>~$}0) with respect to the $^7$Be+\textit{p} channel becomes the major
 one and increases by a factor of 5 as compared with the case of 
\lq\lq white\rq\rq ~  stars (\textit{n$_b$~}=~0, \textit{n$_g$~}=~0). It is obvious that such a tendency is 
connected with an increase of direct proton-nucleon collisions.  
\par
\indent
A further accumulation of statistics on \lq\lq white\rq\rq ~ stars $^8$B$\rightarrow$2He+H is of special interest
 (example is shown in Fig.~2).
 As is known, multiple scattering 
measurements can be used to identify relativistic $^{1,2,3}$H and $^{3,4}$He isotopes. In spite of the fact that these 
measurements are labour demanding, the appropriate efforts can be compensated by the identification of a 3-particle mode
$^8$B$\rightarrow$2$^{3}$He+$^{2}$H   
(threshold of 20 MeV). This possibility is non-trivial because it is connected with a deep rearrangement of the $^8$B cluster 
    structure. The properties of this state near the threshold may turn out to be important for an inverse fusion process
 too. A particular feature of the 2$^{3}$He+$^{2}$H fusion process might consist in a  larger number of vacancies for a neutron 
to be captured in the $^{4}$He cluster formation.
\par
\indent
The conclusions about  the EM dissociation of the  $^8$B and  $^7$Be nuclei \cite{Peresadko06} form the basis for a comparative analysis of 
the  $^9$C nucleus as the next step. The beam of these nuclei is created in the fragmentation of $^{12}$C nuclei of 
an energy of 1.2 GeV, it is used 
to expose emulsion. In all probability, the pattern for $^8$B and  $^7$Be nuclei with the addition of one or,  
respectively, two protons
 must be reproduced for the $^9$C  dissociation. In addition the dissociation $^9$C$\rightarrow$3$^{3}$He
 (threshold of 16 MeV) will become accessible 
for observation. The observation of 3$^{3}$He  population near the threshold would allow one
 to put the ground for an extension 
of the well-known 3$\alpha$ fusion process toward 3$^{3}$He one.
\par
\indent
The emulsion exposed to $^8$B nuclei allowed us to observe events with the total charge of relativistic fragments 
$\Sigma$Z$_{fr}~$~=~6 from 
the mixture of $^{10}$C nuclei produced in the generating target due to the charge exchange process
$^{10}$B$\rightarrow^{10}$C. Their distributions over the charge 
track topology are given in the two lower lines of Table~1. 
Even  restricted statistics points out that the 2He+2H breakup
 accompanied by the conservation  of $\alpha$ particle clusters in 
\lq\lq white\rq\rq ~  stars is more preferable (example is shown in Fig.~3).
 A low $^{8}$Be+2\textit{p} channel threshold equal 
to 3.8 MeV is manifested in such a way.
\par
\indent
The $^{10}$B$\rightarrow^{10}$C charge exchange can be used for further exposures of emulsions with the aim not only to explore the main channel
 of dissociation 2He+2H but also to establish existence of the dissociation mode $^{10}$C$\rightarrow^{4}$He+2$^{3}$He
  (threshold  of 17 MeV). In just the same way as in
 the $^9$C case its discovery can enlarge the picture of the 3He fusion process in nuclear astrophysics. 
To search for  $^8$B$\rightarrow$2He+H    
 and $^{9,10}$C$\rightarrow$3He related to the fragmentation channels it is possible to perform a scanning over the area. This method was already used for
 accelerating the search for the events $^{12}$C$\rightarrow$3He \cite{Belaga95}
 and $^{16}$O$\rightarrow$4He \cite{Andreeva96}, as is discussed below, for $^{9}$Be$\rightarrow$2He.
\par

\section{\label{sec:level3}Fragment Accompaniment of  $^{9}$Be$\rightarrow$2He Dissociation}
\indent
The $^{9}$Be nucleus is a source for the study of the ground and excited states of the $^{8}$Be nucleus. 
Information about the 
generation of a relativistic 2$\alpha$ particle system without the presence of the combinatory  background  of 
other $\alpha$ particles can be utilized in analyzing more complicated N$\alpha$  systems. 
The present-day interest in such systems is inspired by the suggested
search for the $\alpha$ particle Bose--Einstein condensate \cite{Horiuchi03} in which a  ground 0$^+$ and the first excited 
2$^+$ states of the $^{8}$Be nucleus must play the role of condensate basic elements. 
The proof of existence of such a quantum 
state of dilute nuclear matter should be very important in the development of the ideas about nucleosynthesis. 
 The  peripheral dissociation to relativistic N$\alpha$ jets may turn out to be the most convenient tool for searching for it \cite{Andreeva06}. 
\par
\indent
In an emulsion exposed to relativistic $^{9}$Be nuclei 362 events of fragmentation to a narrow pair of 
relativistic He nuclei were analyzed under the assumption of their correspondence to 2$\alpha$ \cite{Artemenkov06,Artemenkov07}. 
A subset of 283 events with \textit{n$_s~$}=~0 is considered below.
Clear apperance  of two peaks 
in the distribution over the invariant mass above the $\alpha$ particle pair mass threshold Q$_{2\alpha}$  was identified.
It was concluded that the 
0$^+$ and 2$^+$   states of $^{8}$Be are revealed in the spectra over Q$_{2\alpha}$.
\par
\indent
The observations of the $^{9}$Be interaction vertices  allows one to separate the population of these states for EM and direct 
nucleon interactions. Table~2 gives the distribution of the $^{9}$Be$\rightarrow$2He events
 in the major intervals over Q$_{2\alpha}$ and the configurations of
 accompanying tracks. The principal feature of the distribution consists in an evident dominance of 144 \lq\lq white\rq\rq ~  
starts (\textit{n$_b$}~=~0, \textit{n$_g$}~=~0) amounting about 60$\%$. Only 27 events (11$\%$) are ascribed  to the stars resulting from $^{9}$Be
 collisions with protons (\textit{n$_b$}~=~0, \textit{n$_g$}~=~1). The ratio of the 
\lq\lq white\rq\rq ~  stars from the states 0$^+$ (Q$_{2\alpha}<$~1 MeV) and 2$^+$ (1$~<~$Q$_{2\alpha}<$~4~MeV) is equal to
 \textit{R$_{0/2}$}~=~3$~\pm~$0.6  and in the case of collisions with target protons (\textit{n$_b$}~=~0, \textit{n$_g$}~=~1) it 
is equal to   \textit{R$_{0/2}$}~=~1~$~\pm$~0.5. Thus, in peripheral fragmentation the production of an $\alpha$ particle pair via the ground
 $^{8}$Be state proceeds more 
intensively than for \textit{n--p} knockout processes. The same conclusion is also valid for the events in which only one target
 nucleus fragment (\textit{n$_b$}~=~1, \textit{n$_g$}~=~0) is revealed and  \textit{R$_{0/2}$}~=~1.5$~\pm~$0.5.
\par
\indent
Following the concept about the $^{9}$Be nucleus as a cluster system $\alpha-$n$-\alpha$ it may be supposed that the ground state of this nucleus 
contains with a noticeable probability a pair of $\alpha$ particle clusters with angular momentum \textit{L}~=~2. The presence 
of a neutron gives the  value for the $^{9}$Be spin 3/2. When the neutron is knocked out by the target 
proton there proceeds either a dispersion of the $\alpha$ particle pair from the D-state or a radiation transition 
to the $^{8}$Be ground state 0$^+$. 
An inverse $^{9}$Be synthesis process might be considered as a radiation transition 
0$^+\rightarrow$~2$^+$ in the
 presence of the neutron. In other words, the $\alpha$ pair goes out from the mass surface with $\gamma$ emission.
 Such a picture is worthy of checking in experiments with  $\gamma$ detection.
\par
\indent
Fig.~4  shows the total transverse momentum distribution transferred to $\alpha$ pairs 
\textit{P$_T^{2\alpha}$} for \lq\lq white\rq\rq ~ stars (\textit{a}) and from a break-ups on protons (\textit{b}). 
The following average values are obtained -   
 \textit{$<$P$_T^{2\alpha}>$}~=138~$\pm$~12 MeV/c (\textit{n$_b$}~=~0, \textit{n$_g$}~=~0) and 194$~\pm~$28 MeV/c (\textit{n$_b$}~=~0,
 \textit{n$_g$}~=~1).
 There is a noticeable difference in the average 
values and the distribution shapes which points to different dynamics of $\alpha$ pair production. 
The $^{9}$Be disintegration on a proton turns out to be a more rigid  reaction mechanism as compared  with disintegration on a Ag, Br. 
An increase in the target fragmentation multiplicity can be expected to result  in an explicit increase in \textit{$<$P$_T^{2\alpha}>$}
 (Table~2). A further growth of the multiplicity corresponding to a larger nucleus overlap leads to a suppression of the 
2He event statistics owing to a destruction of the $^{9}$Be structure destruction. 
\par

\section{\label{sec:level4}Fragment Accompaniment of $^{14}$N and $^{22}$Ne Dissociation}
\indent
The study of the peripheral dissociation of $^{14}$N  nuclei of an energy of 2.1~\textit{A}~GeV in
 the fragment state Z$_{fr}$~=~7 \cite{Shchedrina06,Artemenkov07} has resulted in the conclusion 
about the leading part of the $^{14}$N$\rightarrow$3He channel. Therefore the
 peripheral $^{14}$N dissociation can be used as an effective 
source of 3$\alpha$ systems. The dominant part of the events was shown to be concentrated 
in the interval of the invariant 
3$\alpha$ particle mass over the $^{12}$C mass 7$~<~$Q$_{3\alpha}~<~$20~MeV covering 
the $\alpha$ particle levels  just above the $^{12}$C  dissociation threshold.
Thus, the problems of a few-body nuclear physics near the $\alpha$ emission
 threshold can be explored using detection advantages of 
the relativistic collisions.
\par
\indent
The patteren of the target fragment accompaniment in the $^{14}$N dissociation (Table~3) is
 of the same nature as in the case for $^{9}$Be. In spite 
of some increase in the threshold over Q, the main $\alpha$ particle channel of the systems is 
the EM dissociation on heavy
 nuclei. The \lq\lq white\rq\rq ~  stars dominate while the hydrogen dissociation contribution is not large.
 Within statistical errors the
 ratio of the numbers of events with \textit{n$_b$}~=~0 and 1 for $^{14}$N and $^{9}$Be 
points out that the dissociation mechanisms for 2- and 
3$\alpha$ ensembles are alike.
\par
\indent
As an example of a more complicated system Table~3 gives the description of the accompaniment of 5 events 
$^{22}$Ne for 3.2~\textit{A}~GeV 
selected on the basis of 4100 inelastic interactions \cite{Andreeva06}. The charge-topology distribution of the
 $^{22}$Ne fragmentation in nuclear track emulsion versus target fragment numbers n$_{b}$, n$_{g}$ resulted in Table~4.
In spite of a restricted amount of data one can conclude that 
the generation of 5$\alpha$ particle systems proceeds more preferably  via fragmentation on Ag, Br
(\textit{n$_b$}~=~0, \textit{n$_g$}~=~0). The value of the mean 
transverse momentum transferred to 5$\alpha$ particle systems in these events when normalized to the $\alpha$ particle number
 is located in the region typical for dissociation of light nuclei with a pronounced $\alpha$ clustering.
\par
\indent
The investigations with light nuclei create a methodical basis for the study of exclusively
 complicated systems He-H-\textit{n} for the energy scale relevant for nuclear astrophysics. In this respect, the motivated prospects are associated with 
a detailed analysis of the already observed fragment jets in the events of complete EM dissociation of Au nuclei at 10.6~\textit{A}~GeV and Pb 
nuclei at 160~\textit{A}~GeV.
\par

\section{\label{sec:level5}Conclusions}
\indent
Possessing a record space resolution the nuclear emulsion method keeps unique possibilities in studying the structure 
particularities of light nuclei, first of all, of neutron-deficient nuclei. The presented results of an exclusive study of
 the interactions of relativistic $^{8}$B and $^{9}$Be nuclei in nuclear emulsion lead to the conclusion that the particular features
 of their structure are clearly manifested in peripheral dissociations. In spite of an extraordinarily large distinction
 from the nuclear excitation energy the relativistic scale not only does not impede investigations of nuclear interactions
 in energy scale typical for nuclear astrophysics, but on the contrary gives new methodical advantages. The main advantage 
is the possibility of principle of observing and investigating multi-particle systems. The study of the relativistic 
dissociation of $^{14}$N nucleus to a 3He system confirm this prospect.
\par
\indent
The presented observations can also serve as an illustration of unique prospects of the emulsion method for nuclear 
astrophysics using relativistic nuclei. Providing the three-dimensional observation of dissociation events the nuclear 
emulsion method gives unique possibilities of moving toward more and more complicated nuclear systems generated in peripheral
 dissociations. Therefore this method deserves upgrade, without changes in its basic designation for particle 
detection, with the aim to speed up  the microscope search for rather rare events of peripheral
 dissociation of relativistic nuclei.
\par
\begin{acknowledgments}
\indent The work was supported by the Russian Foundation for Basic Research
 ( Grants 96-159623, 02-02-164-12a,03-02-16134, 03-02-17079 and 04-02-16593 ),
 VEGA 1/9036/02.  Grant from the Agency for Science of the Ministry for Education of the
 Slovak Republic and the Slovak Academy of Sciences, and Grants from the JINR
 Plenipotentiaries of the Republic of Bulgaria, the Slovak Republic, the Czech Republic
 and Romania in the years 2002-2007.\par 
\end{acknowledgments}

\newpage
\begin{figure}
\includegraphics [width=0.85\textwidth ]{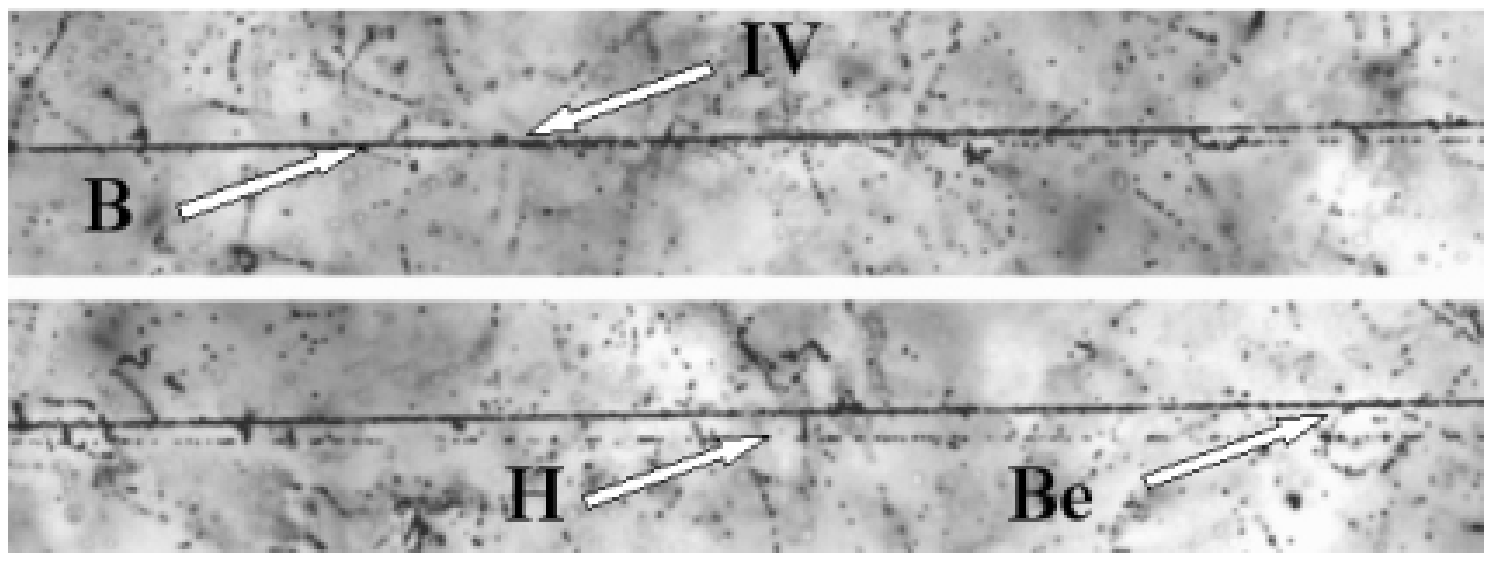}
\caption{Example of  peripheral interaction of a 1.2~\textit{A}~GeV $^{8}$B$\rightarrow^7$Be+p  
 in a nuclear track emulsion (\lq\lq white\rq\rq ~star).
 The interaction vertex (indicated as {\bf IV}) and fragment tracks ({\bf Be} and {\bf H}) in a narrow angular
 cone are seen  on the  microphotograph.
}
 \label{aba:fig1}
\end{figure} 
\begin{figure}
 \includegraphics[width=0.85\textwidth ] {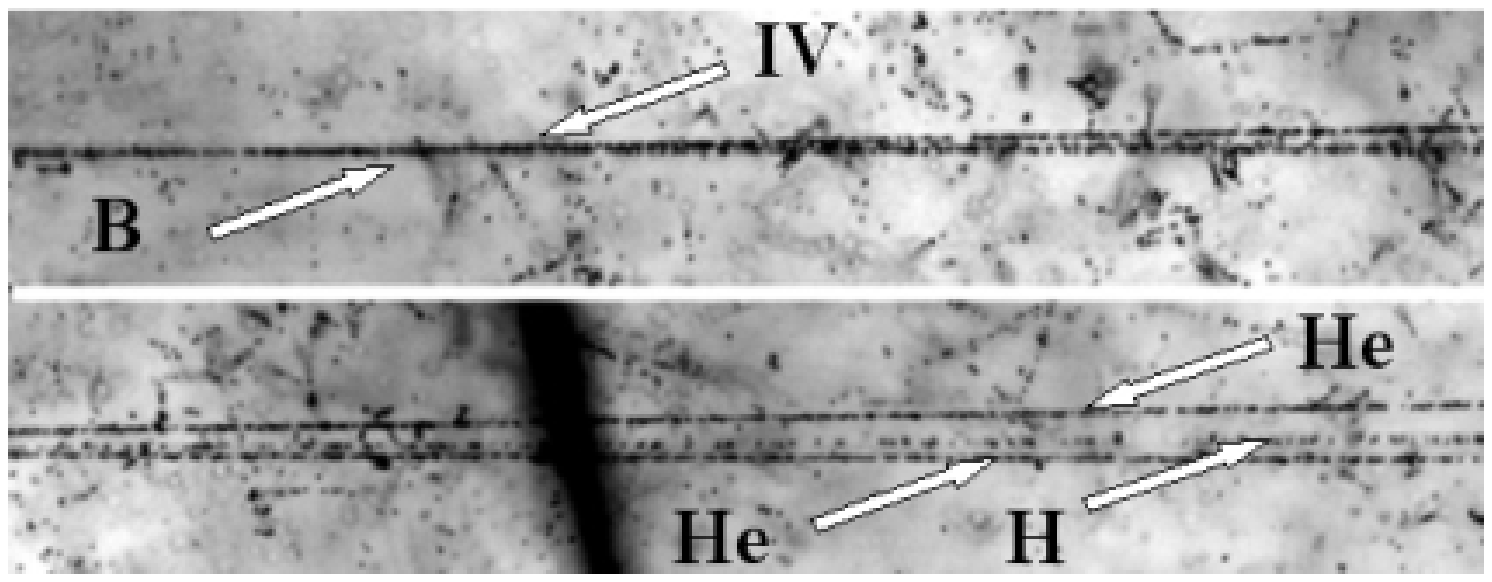}
\caption{Example of peripheral interaction of a 1.2~A~GeV $^{8}$B$\rightarrow$2He+H
 in a nuclear track emulsion (\lq\lq white\rq\rq ~star).
 The interaction vertex (indicated as {\bf IV}) and nuclear fragment tracks ({\bf H} and {\bf He}) in a narrow angular
 cone are seen  on the microphotograph.}
 \label{aba:fig2}
 \end{figure} 
 \begin{figure}
 \includegraphics[width=0.75\textwidth ]{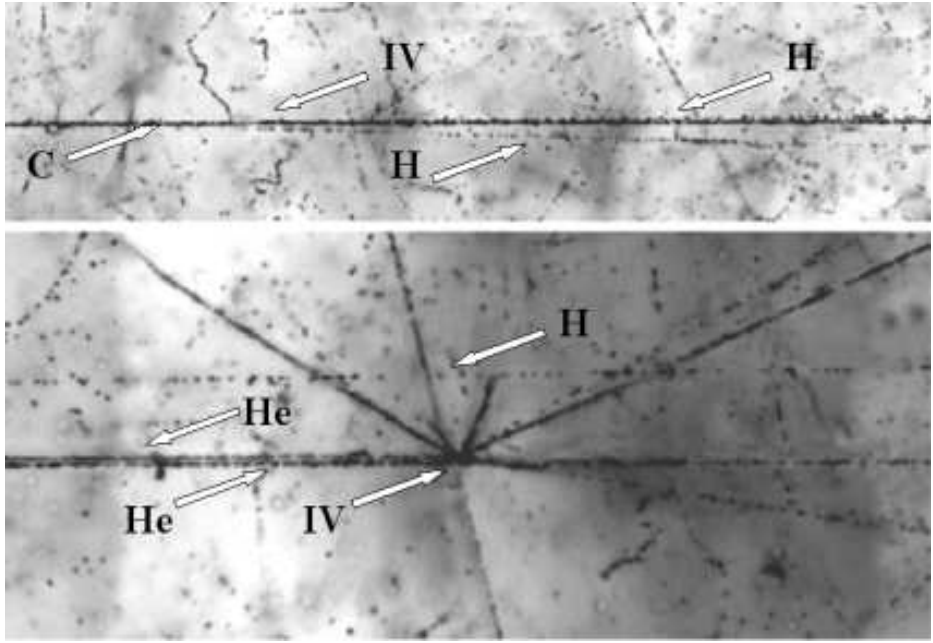}
\caption{Example of peripheral interaction of a 1.2~A~GeV $^{10}$C$\rightarrow$2He+2H  in a nuclear track emulsion (\lq\lq white\rq\rq ~star).
 The interaction vertex (indicated as {\bf IV}) and nuclear fragment tracks ({\bf H} and {\bf He}) in a narrow angular
 cone are seen  on the upper microphotograph.
 Following the direction of the fragment jet, it is possible to distinguish 2 doubly charged 
fragments ($^{8}$Be decay) on  bottom microphotograph. One of He fragments produce secondary star.}
 \label{aba:fig3}
 \end{figure} 
 \begin{figure}
 \includegraphics[width=0.75\textwidth ]{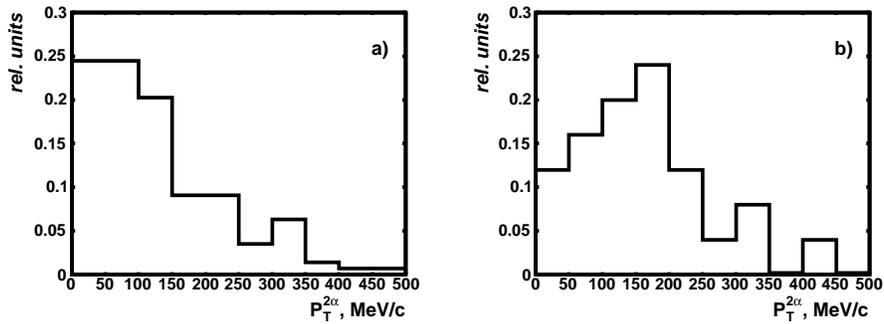}
\caption{The total transverse momentum distribution of $\alpha$ pairs 
P$_T^{2\alpha}$ in 1.2~\textit{A}~GeV $^{9}$Be$\rightarrow$2$\alpha$  for \lq\lq white\rq\rq ~ stars (a) and  break-ups on protons (b).}
 \label{aba:fig4}
\end{figure}
 
\begin{table}
\label{tab:1}
\caption{The distribution of the peripheral interactions with $\Sigma$Z$_{fr}$=5 and
 6 obtained in an emulsion exposed to a $^8$B enriched secondary beam versus target fragment
 numbers n$_b$ and n$_g$.}  
\begin{tabular}{c|c|c|c|c|c|c|c}
\hline\noalign{\smallskip}
~ n$_b$ ~&~ 0 ~&~ 0 ~&~ 1 ~&~ 2 ~&~ 3 ~&~ 4 ~&~ 5 ~\\
~n$_g$~&~ 0 ~&~ 1 ~&~ 0 ~&~ 0 ~&~ 0 ~&~ 0 ~&~ 0 ~\\
\hline\noalign{\smallskip}
 He+3H ~&~ 12 ~&~ 6 ~&~ 3 ~&~ 3 ~&~ 2 ~&~ 3 ~&~ - ~\\
 2He+H ~&~ 14 ~&~ 3 ~&~ 8 ~&~ 2 ~&~ 4 ~&~ - ~&~ 1 ~\\
 Be+H ~&~  25 ~&~ 1 ~&~ 3 ~&~ 3 ~&~ 1 ~&~ - ~&~ - ~\\
 2He+2H ~&~ 3 ~&~ - ~&~ - ~&~ 2 ~&~ - ~&~ 1 ~&~ - ~\\
 He+4H ~&~ - ~&~ 1 ~&~ 3 ~&~ 1 ~&~ 1 ~&~ - ~&~ 1 ~\\
\hline\noalign{\smallskip}
\end{tabular}

\end{table}

\begin{table}
\caption {\label{tab:2}The distribution of the peripheral interactions $^{9}$Be$\rightarrow$2$\alpha$ over intervals
 Q$_{2\alpha}$ versus  target fragment numbers n$_b$ and n$_g$ with corresponding mean values of $\alpha$ pair 
total transverse momentum $<$P$^{2\alpha}_T>$}
\begin{tabular}{c|c|c|c|c|c|c|c}
\hline\noalign{\smallskip}
~ n$_b$ ~&~ 0 ~&~ 0 ~&~ 1 ~&~ 2 ~&~ 3 ~&~ 4 ~&~ 5 ~\\
~n$_g$~&~ 0 ~&~ 1 ~&~ 0 ~&~ 0 ~&~ 0 ~&~ 0 ~&~ 0 ~\\
\hline\noalign{\smallskip}
Q$_{2\alpha}<$1 MeV ~&~ 98 ~&~ 10 ~&~ 21 ~&~ 8 ~&~ 1 ~&~ 3 ~&~ 1 ~\\
 $<$P$^{2\alpha}_T>$, MeV/c ~&~ 133$\pm$16 ~&~ 166$\pm$40 ~&~ 154$\pm$14 ~&~  ~&~  ~&~  ~&~  ~\\
1 MeV$<$Q$_{2\alpha}<$4 MeV ~&~ 33 ~&~ 10 ~&~ 14 ~&~ 3 ~&~ 2 ~&~ 1 ~&~ - ~\\
 $<$P$^{2\alpha}_T>$, MeV/c ~&~ 127$\pm$15 ~&~ 195$\pm$54 ~&~ 178$\pm$23 ~&~  ~&~  ~&~  ~&~  ~\\
4 MeV$<$Q$_{2\alpha}$~&~  13 ~&~ 7 ~&~ 4 ~&~ 2 ~&~ 2 ~&~ 3 ~&~ 1 ~\\
 $<$P$^{2\alpha}_T>$, MeV/c ~&~ 202$\pm$31 ~&~ 232$\pm$42 ~&~ 281$\pm$51 ~&~  ~&~  ~&~  ~&~  ~\\
\hline\noalign{\smallskip}

\end{tabular}
\end{table}
\begin{table}
\caption{\label{tab:3}The distribution of the peripheral interactions $^{14}$N$\rightarrow$3He+H and 
$^{22}$Ne$\rightarrow$5He versus  target fragment numbers n$_b$ and n$_g$ with corresponding mean values of $\alpha$ pair 
total transverse momentum $<$P$^{2\alpha}_T>$}
\begin{tabular}{c|c|c|c|c|c|c|c}
\hline\noalign{\smallskip}
~ n$_b$ ~&~ 0 ~&~ 0 ~&~ 1 ~&~ 2 ~&~ 3 ~&~ 4 ~&~ 5 ~\\
~n$_g$~&~ 0 ~&~ 1 ~&~ 0 ~&~ 0 ~&~ 0 ~&~ 0 ~&~ 0 ~\\
\hline\noalign{\smallskip}
$^{14}$N ~&~ 41 ~&~ 6 ~&~ 23 ~&~ 16 ~&~ 3 ~&~ 2 ~&~ 1 ~\\
 $<$P$^{3\alpha}_T>$, MeV/c ~&~ 222$\pm$21 ~&~ 217$\pm$51 ~&~ 262$\pm$31 ~&~ 378$\pm$54 ~&~  ~&~  ~&~  ~\\
$^{22}$Ne ~&~ 3 ~&~ - ~&~ 1 ~&~ - ~&~ - ~&~ 1 ~&~ - ~\\
 $<$P$^{5\alpha}_T>$, MeV/c ~&~ 518$\pm$85 ~&~ -~&~ - ~&~ - ~&~ - ~&~ - ~&~ - ~\\
 \hline\noalign{\smallskip}
\end{tabular}
\end{table}
\begin{table}[p]
\caption{\label{tab:4}Charge-topology distribution of the
$^{22}$Ne fragmentation in nuclear track emulsion versus target
 fragment numbers n$_{b}$, n$_{g}$ (Channel fraction in percents). }

\bigskip
\begin{tabular}{c|c|c|c|c|c|c}
\hline
~ n$_b$ ~&~ 0 ~&~ 0 ~&~ 1 ~&~ 2 ~&~ 3 ~&~ $\geq$4 ~\\
~n$_g$~&~ 0 ~&~ 1 ~&~ 0 ~&~ 0 ~&~ 0 ~&~ 0 ~\\
\hline
 Ne ~&~ 3 ~&~ 17 ~&~ 118 ~&~ 4 ~&~ 5 ~&~ 4 ~\\
 F+H ~&~ 26~(19,5) ~&~ 9~(15,0) ~&~ 13~(44,8) ~&~ 2 ~&~ - ~&~ 1 ~\\
 O+He ~&~ 54~(40,6) ~&~ 19~(31,7) ~&~ 2~(6,9) ~&~ - ~&~ 1 ~&~ 1 ~\\
 O+2H ~&~ 12~(9,0) ~&~ 7~(11,7) ~&~ - ~&~ - ~&~ - ~&~ - ~\\
 N+He+H ~&~ 12~(9,0) ~&~ 7(11,7) ~&~ 4~(13,8) ~&~ 1 ~&~ - ~&~ - ~\\
 N+3H ~&~ 3~(2,3) ~&~ 3~(5,0) ~&~ - ~&~ - ~&~ - ~&~ - ~\\
 C+2He ~&~ 5~(3,8) ~&~ 3~(5,0) ~&~ 3~(10,3) ~&~ 1 ~&~ - ~&~ - ~\\
 C+2He+2H ~&~ 5~(3,8) ~&~ 3~(5,0) ~&~ 3~(10,3) ~&~ - ~&~ - ~&~ - ~\\
 C+4H ~&~ 2~(1,5) ~&~ - ~&~ - ~&~ - ~&~ - ~&~ - ~\\
 B+Li+He ~&~ 1~(0,8) ~&~ - ~&~ - ~&~ - ~&~ - ~&~ - ~\\
 B+2He+H ~&~ 2~(1,5) ~&~ 1~(1,7) ~&~ - ~&~ - ~&~ -~&~ - ~\\
 B+He+3H ~&~ 2~(1,5) ~&~ 1~(1,7) ~&~ - ~&~ - ~&~ - ~&~ - ~\\
 B+5H ~&~ 1~(0,8) ~&~ - ~&~ 1~(3,4) ~&~ - ~&~ - ~&~ - ~\\
 2Be+2H ~&~ - ~&~1~(1,7) ~&~ - ~&~ - ~&~ - ~&~ - ~\\
 Be+Li+3H ~&~ 1~(0,8) ~&~ - ~&~ - ~&~ - ~&~ - ~&~ - ~\\
 Be+3He ~&~ 2~(1,5) ~&~ - ~&~ -~&~ - ~&~ - ~&~ - ~\\
 Be+2He+2H ~&~ - ~&~ - ~&~ - ~&~ - ~&~ 1~&~ - ~\\
 Be+He+4H ~&~ 1~(0,8) ~&~ - ~&~ - ~&~ - ~&~ - ~&~ - ~\\
 Li+3He+H ~&~ - ~&~ 1~(1,7) ~&~ - ~&~ - ~&~ - ~&~ - ~\\
 5He ~&~ 3~(2,3) ~&~ - ~&~ 1~(3,4) ~&~ - ~&~ - ~&~ 1 ~\\
 4He+2H ~&~ 1~(0,8) ~&~ 5~(8,3) ~&~ 2~(6,9) ~&~ - ~&~ - ~&~ - ~\\
 3He+4H ~&~ - ~&~ - ~&~ -~&~ 2 ~&~ - ~&~ - ~\\
\hline
Sum~&~ 136 ~&~ 77 ~&~ 147 ~&~ 10 ~&~ 7 ~&~ 7 ~\\
[1mm]
\hline
\end{tabular}
\end{table}

\newpage

\end{document}